\documentclass[12pt,longnamesfirst,preprint]{aastex}

\newcommand{\myemail}{yokogawa@nro.nao.ac.jp}

\slugcomment{Accepted for Publication in the Astrophysical Journal}

\shorttitle{Yokogawa et al.}
\shortauthors{CS $J=2-1$ and $J=3-2$ imaging of L1551 NE}

\begin{document}


\title{
High Angular Resolution, Sensitive CS $J=2-1$ and $J=3-2$ Imaging of the
Protostar L1551 NE:\\  Evidence for Outflow-Triggered Star Formation
?\altaffilmark{1}}


\author{
Sozo Yokogawa\altaffilmark{2,3},Yoshimi Kitamura\altaffilmark{4},
Munetake Momose\altaffilmark{5}, and Ryohei Kawabe\altaffilmark{3}} 


\altaffiltext{1}{
Based on the observations made at the Nobeyama Radio Observatory (NRO),
which is a branch of the National Astronomical Observatory, an
interuniversity research institute operated by the Ministry of
Education, Science, Sports, Culture, and Technology.}
\altaffiltext{2}{
Department of Astronomical Science, The Graduate University for Advanced
Studies, Osawa 2-21-1, Mitaka, Tokyo, 181-8588, Japan} 
\altaffiltext{3}{National Astronomical Observatory of Japan, Osawa 2-21-1,
Mitaka, Tokyo, 181-8588, Japan; {\myemail}, kawabe@nro.nao.ac.jp}
\altaffiltext{4}{Institute of Space and Astronautical Science,
Yoshinodai 3-1-1, Sagamihara, Kanagawa, 229-8510, Japan;
kitamura@pub.isas.ac.jp}  
\altaffiltext{5}{Institute of Astrophysics \& Planetary Sciences, 
Ibaraki University, Bunkyo 2-1-1, Mito, Ibaraki, 310-8512, Japan;
momose@mx.ibaraki.ac.jp} 


\begin{abstract}
 High angular resolution and sensitive aperture synthesis observations of   
 CS ($J=2-1$) and CS ($J=3-2$) emissions toward L1551 NE, the second
 brightest protostar in the Taurus Molecular Cloud, made with the 
 Nobeyama Millimeter Array are presented. 
 L1551 NE is categorized as a class 0 object deeply embedded in the
 red-shifted outflow lobe of L1551 IRS 5. 
 Previous studies of the L1551 NE region in CS emission revealed the
 presence of shell-like components open toward L1551 IRS 5, which seem to trace
 low-velocity shocks in the swept-up shell driven by the outflow from
 L1551 IRS 5.    
 In this study, significant CS emission around L1551 NE was detected at the
 eastern tip of the swept-up shell from $V_{\rm{lsr}}$ = 5.3 km s$^{-1}$
 to 10.1 km s$^{-1}$, and the total mass of the dense  gas is estimated
 to be 0.18 $\pm$ 0.02 $M_\odot$. Additionally, the following new
 structures were successfully revealed:  
 a compact disklike component with a size of $\approx$ 1000 AU just at
 L1551 NE, an arc-shaped structure around L1551 NE, open toward L1551 NE,
 with a size of $\sim 5000$ AU, i.e., a bow shock, and a distinct
 velocity gradient of the dense gas, i.e., deceleration along the
 outflow axis of L1551 IRS 5. 
 These features suggest that the CS emission traces the post-shocked
 region where the dense gas associated with L1551 NE and the swept-up
 shell of the outflow from L1551 IRS 5 interact. 
 Since the age of L1551 NE is comparable to the timescale of the interaction,
 it is plausible that the formation of L1551 NE was induced by the outflow
 impact. The compact structure of L1551 NE with a tiny envelope was also
 revealed, suggesting that the outer envelope of L1551 NE has been blown
 off by the outflow from L1551 IRS 5.  
\end{abstract}


 \section{Introduction}
Star formation is often affected by various external disturbances
due to active phenomena such as expansion of H II regions, UV
radiation from massive stars, and shock waves from supernova remnants.    
In contrast, the Taurus Molecular Cloud
(TMC), the most intensively studied low-mass star-forming region lying at
a distance of 140 pc ~\citep{Elias78}, has been regarded as a site
where star formation occurs spontaneously: the formation
of each star progresses in an isolated environment.  
This is supported by the following two facts: 1) Although the TMC is located
near the center of the Cas-Tau OB association, the association is old
($\sim$25 Myr) and has already been dispersed into a wide area of 90$^\circ$
$\times$ 40$^\circ$ ~\citep{1991psfe.conf..125B}. 
2) Since the number density of the
YSOs in the association is quite low, $10-20$ stars pc$^{-3}$
~\citep{1993AJ....105.1927G}, the neighboring systems cannot
gravitationally interact with  
each other during star formation. In the L1551 dark cloud, however, the
protostars such as L1551 IRS 5 (hereinafter IRS 5), L1551 NE
(hereinafter NE), HL Tau, and HH 30
might be mutually affected through their outflows. IRS 5, the most
luminous protostar ($L_{\rm{bol}}$ $\approx$ 30$L_\odot$) in the TMC,
has powerful outflows, and NE is located just in its red-shifted outflow
lobe. It is therefore likely that the environment around NE is strongly
affected by the outflows ~\citep{1994ApJ...425L..93B}. Furthermore,
both HL Tau and HH 30 have optical jets deeply penetrating the L1551 region,
which would disturb the entire cloud: recent studies have revealed that
dozens of parsec-scale HH flows driven by low-mass young stars are
thought to disrupt surrounding molecular cloud 
cores and drive supersonic turbulence ~\citep{1997AJ....114.2708R,
2001ARA&A..39..403R}. Consequently, even in the TMC, outflows can play
important roles in neighboring star formation. 

L1551 NE, which was discovered by Emerson et al. (1984), is the second
brightest protostar ($L_{\rm{bol}}$ $\approx$ 6$L_\odot$) in the TMC, 
and is categorized as a class 0 object (e.g., Barsony \& Chandler 1993).
Recent high angular resolution observations at $\lambda$ =
3.6 cm have revealed that NE is a close binary of sources A and B with
a separation of 0$\farcs$5 at P.A. of 297$^\circ$
~\citep{2002AJ....124.1045R}.  
The sources are likely to drive Harbig-Haro objects HH 28, HH 29, and HH 454 as
well as the well collimated infrared [Fe II] jet
~\citep{1999AJ....118..972D, 2000AJ....120.1449R}.
In addition, the interaction features between NE and the outflow lobe of
IRS 5 have been revealed: high-velocity CS emission was detected around NE,
and the distribution of the emission has a clumpy and shell-like
structure open toward IRS 5, suggesting that the swept-up shell of the
outflow from IRS 5 has impacted on NE ~\citep{1995ApJ...446..234P}.    
The detailed spatial and velocity structures of the swept-up shell,
however, are still unknown owing to the insufficient angular resolutions
and sensitivities. To investigate the detailed structures of the dense
gas around NE and to reveal the relationship between NE and the
red-shifted outflow lobe of IRS 5, we made aperture synthesis
observations of NE in CS ($J=2-1$) and CS ($J=3-2$) emissions, which
trace high-density and shocked gas, with higher angular resolution and
better sensitivity.
 
\section{Observations}
\subsection{CS ($J=2-1$) and CS ($J=3-2$) emissions with the
Nobeyama Millimeter Array}
Aperture synthesis observations of NE were carried out in CS ($J=2-1$)
(97.980968 GHz) and CS ($J=3-2$) (146.969049 GHz) with the
Nobeyama Millimeter Array (NMA), which consists of six 10 m antennas,
during a period from 1999 November to 2002 February. The CS images were
obtained using all the array configurations, D, C, and AB, whose ranges of
the projected baseline lengths were $3-27$, $6-55$, and $15-110$ k$\lambda$ at
98 GHz, and $5-40$, $10-80$, and $25-160$ k$\lambda$ at 147 GHz, respectively. 
The system noise temperatures of SIS receivers in DSB mode were about
150 K at 98 GHz and 200 K at 147 GHz toward the zenith. We used a 1024
channel FX spectrocorrelator with a total bandwidth of 32 MHz, resulting
in velocity resolutions 
of 0.095 km s$^{-1}$ at 98 GHz and 0.063 km s$^{-1}$ at 147 GHz.   
The center of the field was set on the NE position of ($\alpha_{2000}$,
$\delta_{2000}$) = (4$^h$ 31$^m$ 44$^s$.42, +18$^\circ$ 08$\arcmin$
32$\farcs$3). The FWHM sizes of the primary beam were 70$\arcsec$ for 98
GHz and 51$\arcsec$ for 147 GHz. Since the 
minimum baseline lengths at 98 GHz and 147 GHz were 3 k$\lambda$ and 5
k$\lambda$, respectively, our observations were insensitive to
structures extending more than 69$\arcsec$ (10$^4$ AU at 140 pc) at 98
GHz and 41$\arcsec$ (6 $\times$ 10$^3$ AU) at 147 GHz. 
The response across the observed passband for each sideband was determined
from $30-40$ minutes observations of 3C454.3 and 3C279. Gain calibrators
0446+112, 0507+179, 0528+134 were observed every $10-20$ minutes. The flux
density of each calibrator was derived from observations of Uranus.
The overall uncertainty in the flux calibration was about 10 \%.
After these calibrations, only the data taken under good weather
conditions were used in imaging. Using the AIPS package
developed at the NRAO, we CLEANed maps by natural weighting with no
taper in the UV plane. 

The continuum data were also obtained with the
digital spectral correlator UWBC ~\citep{2000PASJ...52..393O}, which has 128
frequency channels and a 1024 MHz bandwidth per baseline. Visibility data
of 100 GHz ($\lambda$ = 3 mm) continuum emission in both the lower ($98
\pm 0.512$ GHz) and upper ($110 \pm 0.512$ GHz) sidebands were obtained
simultaneously with a phase-switching technique. The data of 150 GHz
($\lambda$ = 2 mm) continuum emission were obtained in the lower ($135
\pm 0.512$ GHz) and upper ($147 \pm 0.512$ GHz) sidebands. 
More details of the observational parameters are summarized in
Table \ref{tbl:tbl1}.   

\placetable{tbl:tbl1}

\subsection{CS ($J=2-1$) emission with the Nobeyama 45 m telescope}
Mapping observations of the CS ($J=2-1$) emission were carried out
in 2001 February with the Nobeyama 45 m telescope. The beam size (HPBW)
was 16$\arcsec$ at the frequency of CS ($J=2-1$). The main-beam
efficiency, $\eta_{\rm{MB}}$, was about 0.5. We used a cooled SIS
receiver with a single sideband filter. 
The typical system noise temperature was 350 K (in SSB) at an elevation
of 70$^\circ$. At the backend, an acousto-optic-spectrometer (AOS)
which has 2048 channels with a 40 MHz bandwidth was employed, and the
frequency resolution was 37 kHz corresponding to 0.11 km s$^{-1}$ for CS
($J=2-1$). 
A 2$\arcmin$ $\times$ 2$\arcmin$ area whose center is the NE position
was mapped with a grid spacing of 10$\arcsec$.
The typical noise level in the map was 0.2 K in $T^{*}_{\rm{A}}$.  

\section{Results}
\subsection{3 mm and 2  mm continuum emission}
\begin{figure}
\epsscale{0.7}
\plotone{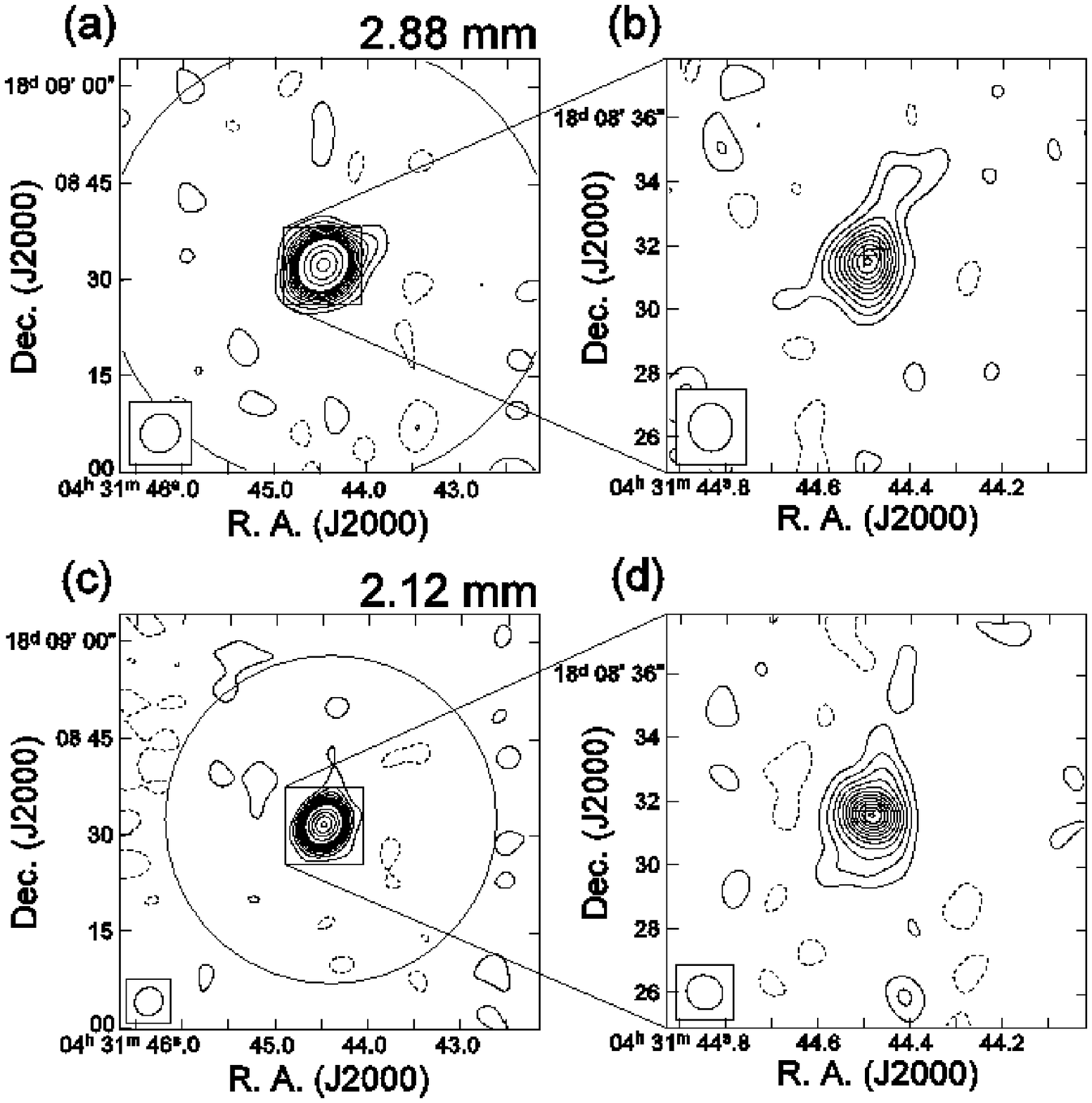}
\caption{
 (a), (b) Lowest and highest angular resolution images with the D and AB
 configurations at $\lambda$ = 2.88 mm, respectively. 
 (c), (d) Lowest and highest angular resolution images with the D and AB
 configurations at $\lambda$ $= 2.12$ mm, respectively. 
 The contour lines of these maps start at $\pm$2.0~$\sigma$ levels with
 intervals of 2.0~$\sigma$ until 20~$\sigma$, and the intervals become
 10~$\sigma$ above 20~$\sigma$. The negative levels are indicated by
 broken lines.  
 The two crosses in the (b) \& (d) maps indicate the binary positions
 determined at $\lambda$ $= 3.6$ cm with the
 VLA~\citep{2002AJ....124.1045R}.    
 The rms noise levels (1~$\sigma$) of the (a), (b), (c), and (d) maps
 are 1.4, 2.0, 2.6, and 2.6 mJy beam$^{-1}$, respectively. The open
 ellipse in the bottom left corner of each map is the synthesized beam
 size (HPBW).   
\label{fig:fig1}}
\end{figure}

We have obtained $\lambda$ = 3 mm and 2 mm continuum maps of NE with
$6\arcsec - 1\arcsec$ resolutions. Continuum emission at these
wavelengths is thought to mainly come from thermal radiation of dust
particles. Maps with the
lowest and highest angular resolutions are shown in Figure
\ref{fig:fig1}.   
To obtain higher signal-to-noise ratios (S/N) in the 3 mm and 2 mm
images, the data of both the lower and upper sidebands were 
combined into final images. Therefore, the center frequency of the 3 mm
continuum images is 104 GHz ($\lambda$ = 2.88 mm), and that of the 2 mm
continuum images is 141 GHz ($\lambda$ = 2.12 mm). 
The total flux densities of the continuum emission are 92.5 $\pm$ 9.4
mJy at 104 GHz and 202 $\pm$ 22 mJy at 141 GHz with the D
configuration, and are 68.3 $\pm$ 7.1 mJy at 104 GHz and 130 $\pm$ 13
mJy at 141 GHz with the AB configuration (see Table
\ref{tbl:tbl2}). 
Both the images in the AB configuration show disklike structures
elongating along the direction at P.A. = 159$^\circ$ $\pm$ 13$^\circ$ 
with the mean size of 1$\farcs$39 $\times$ 1$\farcs$00 (FWHM) (200 AU
$\times$ 140 AU), which is almost perpendicular to both the [Fe II] jet
~\citep[P.A. $\approx$ 243$^\circ$]{2000AJ....120.1449R} and the
HH jets HH 28, HH 29, and HH 454 ~\citep[P.A. $\approx$
242$^\circ$]{1999AJ....118..972D}.   
The peak positions of the continuum emission are 
($\alpha_{2000}$, $\delta_{2000}$) = (04$^h$ 31$^m$ 44$^s$.487,
+18$^\circ$ 08$\arcmin$ 31$\farcs$57) at 3 mm, and ($\alpha_{2000}$,
$\delta_{2000}$) = (04$^h$ 31$^m$ 44$^s$.482, +18$^\circ$ 08$\arcmin$
31$\farcs$58) at 2 mm. 
Since the typical positional error, which includes the effect of atmospheric
seeing, is 0$\farcs$3, both the peak positions are the same within the
errors. 
We regard these positions as the source position of NE.  

\placefigure{fig:fig1}
\placetable{tbl:tbl2}

Recent VLA observations at $\lambda$ = 3.6 cm revealed that NE is a
close binary system with a 0$\farcs$5 separation, 
corresponding to 70 AU at 140 pc. The stellar positions are
($\alpha_{2000}$, $\delta_{2000}$) = (04$^h$ 31$^m$ 44$^s$.497,
+18$^\circ$ 08$\arcmin$ 31$\farcs$67) for NE-A, and ($\alpha_{2000}$,
$\delta_{2000}$) = (04$^h$ 31$^m$ 44$^s$.465, +18$^\circ$ 08$\arcmin$
31$\farcs$88) for NE-B ~\citep{2002AJ....124.1045R}, which are shown as
crosses in Figure \ref{fig:fig1}.  
Since both the binary components are surrounded by the disklike
structure we revealed, this corresponds to a circumbinary
disk around the stars.  
The peak positions of the 3 mm and 2 mm continuum emission, however, are
almost consistent with the position of NE-A which probably ejects the
[Fe II] jet ~\citep{2000AJ....120.1449R}, suggesting that the dust
is more concentrated at NE-A than NE-B. In Figure \ref{fig:fig1}, one
cannot identify any inner hole, which could be generated by tidal
force from the binary system.
This is probably because NE is still in an active accretion stage and
the accreting matter prevents the growth of the inner hole.
To spatially resolve such a detailed structure of the circumbinary disk, 
extremely high angular resolution observations are required, and will be
done with the Atacama Large Millimeter/ submillimeter Array (ALMA).  
The presence of another companion located to the 1$\farcs$4 south of
NE-A and B is suggested by Moriarty-Schieven et al. (2000), but we did not
detect such a companion in both the 3 mm and 2 mm images in spite of our
sufficient angular resolutions. 

Assuming that the dust continuum emission around NE is optically thin at
the millimeter  wavelengths, and the dust temperature, $T_{\rm{dust}}$,
is uniform, we can estimate the mass of the disklike structure by 
\begin{equation}
M_{\rm{NE}} = \frac{F_{\nu} d{^2}}{{\kappa}_{\nu} B_{\nu}(T_{\rm{dust}})},
\end{equation}
where $\kappa_\nu$ is the dust mass absorption coefficient which is
usually given by $0.1(\nu/10^{12}~\rm{Hz})^{\beta}~$cm$^2$~g$^{-1}$
~\citep{1990AJ.....99..924B}, $B_{\nu}(T)$ is the Planck function, $F_{\nu}$
is the total flux density of the continuum emission, and $d$ is the
distance to NE (140 pc). The power-law index of $\kappa_\nu$, $\beta$,
is derived as 0.70 from our results and the previous
interferometric studies at 87 GHz and 230 GHz
~\citep{2001ApJ...547..840S, 2000ApJ...533L.143M}.    
With $T_{\rm{dust}}$ = 42 K ~\citep{1994ApJ...436..800M}, a mass of
0.032 $\pm$ 0.003 $M_{\odot}$ is obtained.  

\subsection{CS emission} 
\subsubsection{CS ($J=2-1$) and CS ($J=3-2$) emissions with the
Nobeyama Millimeter Array} 
\begin{figure}
\epsscale{1.0}
\plotone{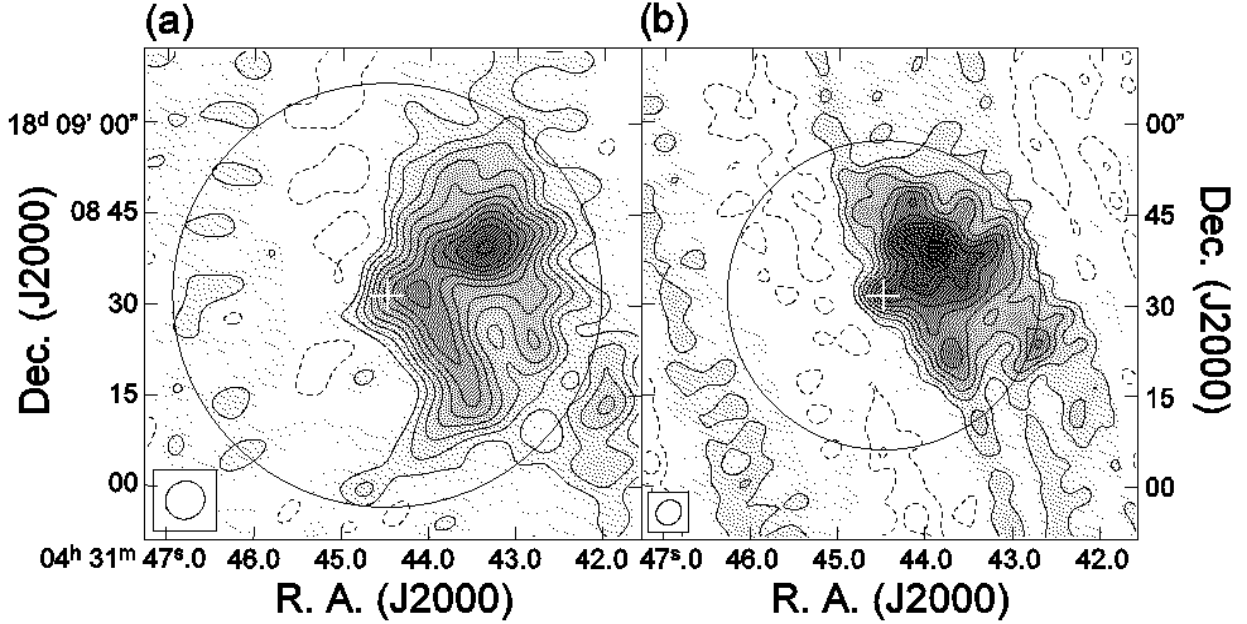}
\caption{
 (a) Total intensity map of CS ($J=2-1$) emission integrated
 from $V_{\rm{lsr}} = 5.32$ km s$^{-1}$ to 10.11 km s$^{-1}$. The
 contour lines start at $\pm$2.0~$\sigma$ levels with intervals of
 2.0~$\sigma$ until 20~$\sigma$, and the intervals become 4.0~$\sigma$
 above 20~$\sigma$. The negative levels are indicated by broken lines.   
 The rms noise level (1~$\sigma$) is 0.025 Jy beam$^{-1}$, and the peak flux
 density is 0.61 Jy beam$^{-1}$. 
 (b) Total intensity map of CS ($J=3-2$) emission 
 integrated from 5.43 km s$^{-1}$ to 10.04 km s$^{-1}$. The contour lines
 are indicated in the same manner as in (a). The rms noise level
 (1~$\sigma$) is  0.040 Jy beam$^{-1}$, and the peak flux density is 0.86
 Jy beam$^{-1}$. 
 The circles on both the maps are the FWHM sizes of the primary beam.
 The open ellipse in the bottom-left corner in each map is the
 synthesized beam (HPBW).  
 The central cross in each map is the position of NE.\label{fig:fig2}
 }
\end{figure}

The emission above the 3~$\sigma$ level is detected within the velocity
range of $V_{\rm{lsr}} = 5.32 - 10.11$ km s$^{-1}$ for CS ($J=2-1$) and
$V_{\rm{lsr}} = 5.43 - 10.04$ km s$^{-1}$ for CS ($J=3-2$). The total
intensity maps integrated over these ranges are shown in Figure
\ref{fig:fig2}.  
Both the images have revealed extended and clumpy components on the west
side of NE. The brightest peak is seen to the $15\arcsec - 20\arcsec$
northwest of NE. Another bright component is seen in the vicinity of NE. The
elongated emission from the NE position to the southwest, which looks
like a ridge with 20$\arcsec$ in length, is seen in the CS ($J=2-1$)
map, and a corresponding southwest peak is seen in the CS ($J=3-2$)
map. No distinct emission, however, is seen on the east side of NE. 

\begin{figure}
\epsscale{0.7}
\plotone{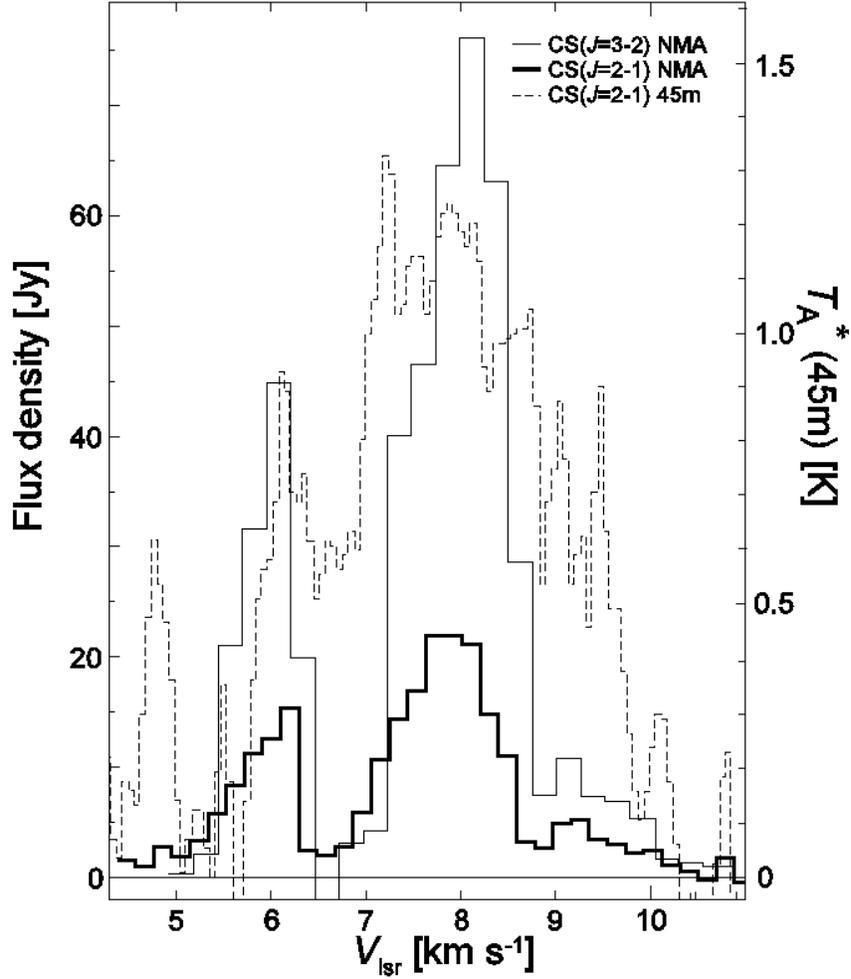}
\caption{
 CS ($J=2-1$) and ($J=3-2$) line profiles taken with the NMA and
 the Nobeyama 45 m telescope. These profiles are integrated over a
 square region of 50$\arcsec$ $\times$ 50$\arcsec$ around NE. 
 The solid lines represent the profiles taken by the NMA which is
 applied the correction for the primary beam attenuation:  
 thick line is the profile of CS ($J=2-1$) and thin one is that of CS
 ($J=3-2$) in unit of Jy. 
 The dashed line represents the CS ($J=2-1$) profile taken by the
 45 m telescope in unit of $T_A^*$. The $T_A^*$ scale of the right axis
 is equivalent to the Jy scale of the left one. 
 \label{fig:fig3}
 }
\end{figure}

\placefigure{fig:fig2}
\placefigure{fig:fig3}

\begin{figure}
\epsscale{0.8}
\plotone{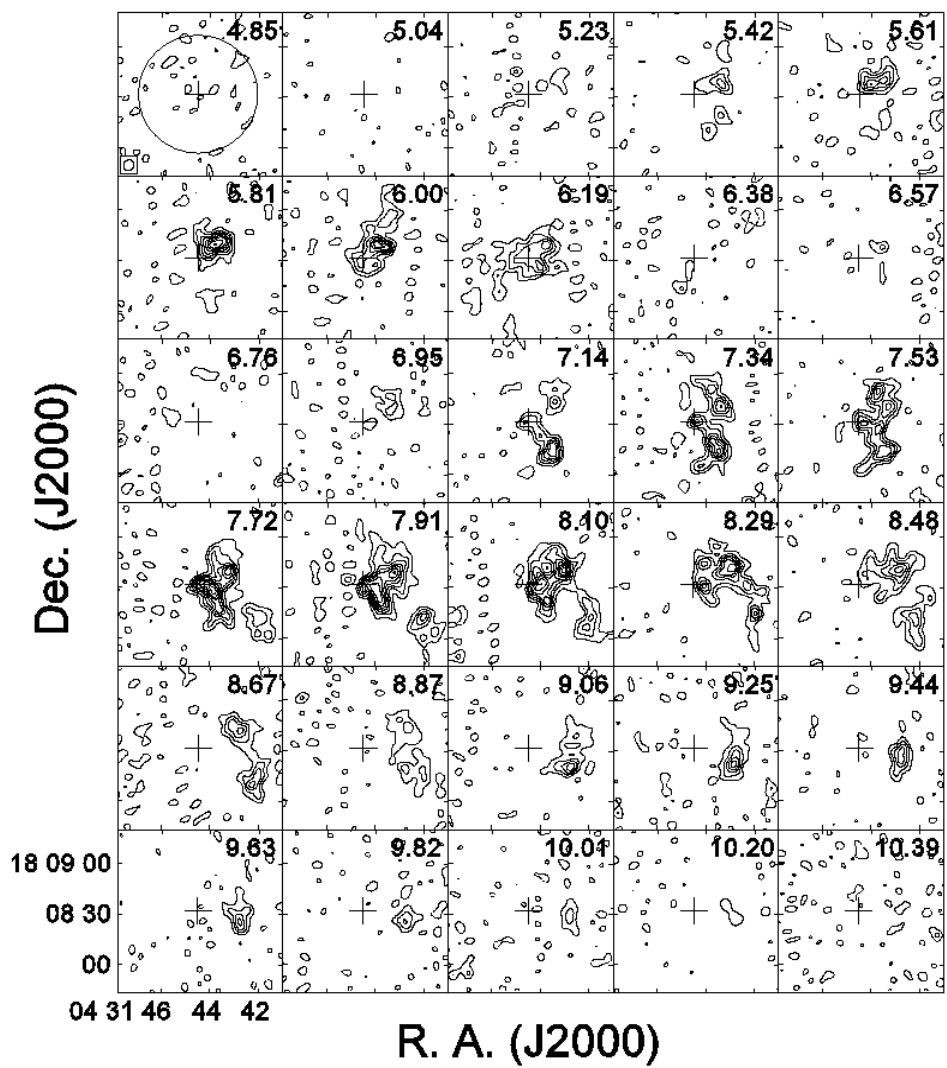}
\caption{
 Velocity channel maps of CS ($J=2-1$) emission. The
 central LSR velocity is denoted in unit of km s$^{-1}$ at the upper
 right corner of each panel. The contour intervals are 2.0~$\sigma$,
 starting at $\pm$2.0~$\sigma$ levels with 1~$\sigma = 0.12$ Jy beam$^{-1}$. 
 The negative levels are indicated by broken lines. The central cross in
 each map exhibits the position of NE. The open ellipse in the
 bottom-left corner of the top-left panel is the synthesized beam
 (HPBW).\label{fig:fig4}} 
\end{figure}

Figure \ref{fig:fig3} shows the CS ($J=2-1$) and CS ($J=3-2$) line
profiles integrated over a 50$\arcsec$ square around NE as well as the
CS ($J=2-1$) line profile in the same region obtained with the 45 m
telescope, which will be described in detail in \S~\ref{res2}.  
Both the NMA profiles can be decomposed into three components:
a blue-shifted component smaller than $\sim 6.5$ km s$^{-1}$ (hereafter
low-velocity blue: LVB), a red-shifted component in $7 - 8.5$ km s$^{-1}$
(hereafter low-velocity red: LVR), and a highly red-shifted wing component
greater than $\sim$ 8.5 km s$^{-1}$ (hereafter high-velocity red: HVR). 
The emission around the systemic velocity of 6.7 km s$^{-1}$, which was
determined by the optically thin H$^{13}$CO$^{+}$ line emission 
~\citep{2001ApJ...547..840S}, is not detected owing to
self-absorption and resolving out of extended components caused by the
lack of short spacing data in our observations. 
The three components can also be identified in channel maps (Figures
\ref{fig:fig4} \& \ref{fig:fig6}) and a position-velocity (P-V)
diagram along the outflow axis from IRS 5 (Figure \ref{fig:fig8}a), which
will be described later.  

Figure \ref{fig:fig4} shows channel maps of CS ($J=2-1$)
with a velocity resolution of 0.19 km s$^{-1}$.
From $V_{\rm{lsr}} = 5.42$ km s$^{-1}$ to 6.19 km s$^{-1}$, the CS gas
is mainly distributed to the northwest of NE. In just the vicinity of
NE, however, there
is no emission in the range of $V_{\rm{lsr}} = 5.42 - 5.81$ km s$^{-1}$ and
is only weak extended emission in $V_{\rm{lsr}} = 6.00$ and 6.19 km s$^{-1}$.  
No significant emission is detected in the range of $6.38 - 6.95$ km
s$^{-1}$, including the systemic velocity, mainly owing to the resolving out. 
In the following channels from $V_{\rm{lsr}} = 7.14$ km s$^{-1}$ to 7.53 km
s$^{-1}$, a compact component is detected just at the NE
position. The beam deconvolved size of the component is
11$\farcs$0 $\times$ 6$\farcs$5 (FWHM) (1540 $\times$ 910 AU) at P.A. of
60$^\circ$. This component possibly comes from the disk/envelope system
around NE (hereafter the NE component). 
In these channels, other strong peaks are seen to the 
northwest and southwest of NE, and an inclined ``W'' shaped structure,
which contains all the peaks, can be 
recognized. From $V_{\rm{lsr}} = 7.72$ km s$^{-1}$ to 8.29 km s$^{-1}$,
the NE component and a peak located $\sim$ 20$\arcsec$
northwest of NE are distinct.  
A new weak component $\sim$ 40$\arcsec$ southwest of NE appears beyond
$V_{\rm{lsr}} = 7.72$ km s$^{-1}$.  
In the channels from $V_{\rm{lsr}} = 8.48$ km s$^{-1}$ to 10.01 km
s$^{-1}$, the southwest and northwest components gradually merge
into a west component which is apparent beyond $V_{\rm{lsr}} = 9.25$
km s$^{-1}$.   

\begin{figure}
\epsscale{1.0}
\plotone{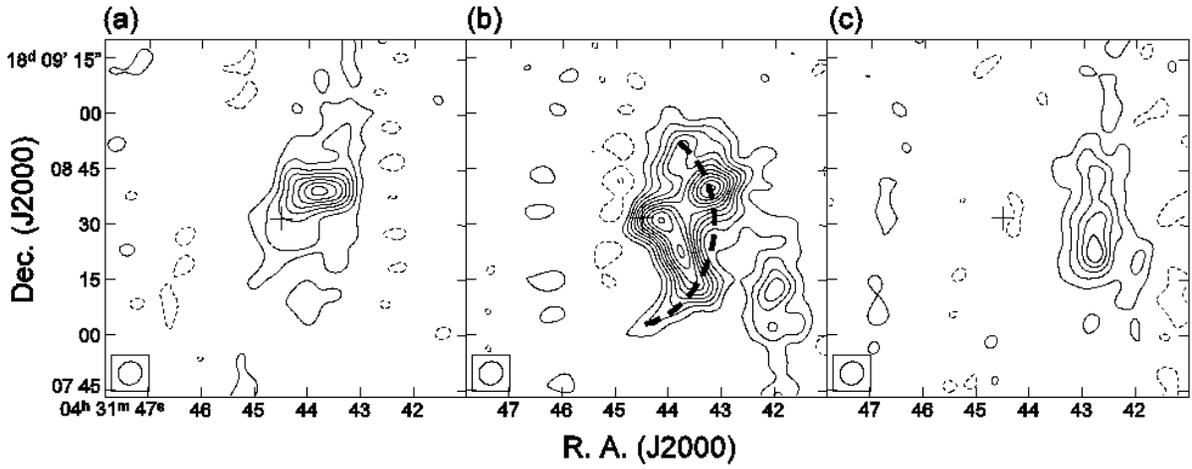}
\caption{
 CS ($J=2-1$) maps for (a) the low-velocity blue (LVB) component of
 $V_{\rm{lsr}} = 5.32 - 6.47$ km s$^{-1}$, (b) the low-velocity
 red (LVR) component of $V_{\rm{lsr}} = 7.04 - 8.57$ km s$^{-1}$
 , and (c) the high-velocity red (HVR) component of $V_{\rm{lsr}} = 8.57 -
 10.11$ km s$^{-1}$. The contour lines of each map are
 written in the same manner as in Figure \ref{fig:fig4}.
 The rms noise levels are 62 mJy beam$^{-1}$, 54 mJy
 beam$^{-1}$, and 57 mJy beam$^{-1}$ for (a), (b), and (c),
 respectively. The central cross is the position of NE.
 The dashed line in (b) delineates the newly found small-scale
 arc open toward NE. \label{fig:fig5}}
\end{figure}

\begin{figure}
\epsscale{0.8}
\plotone{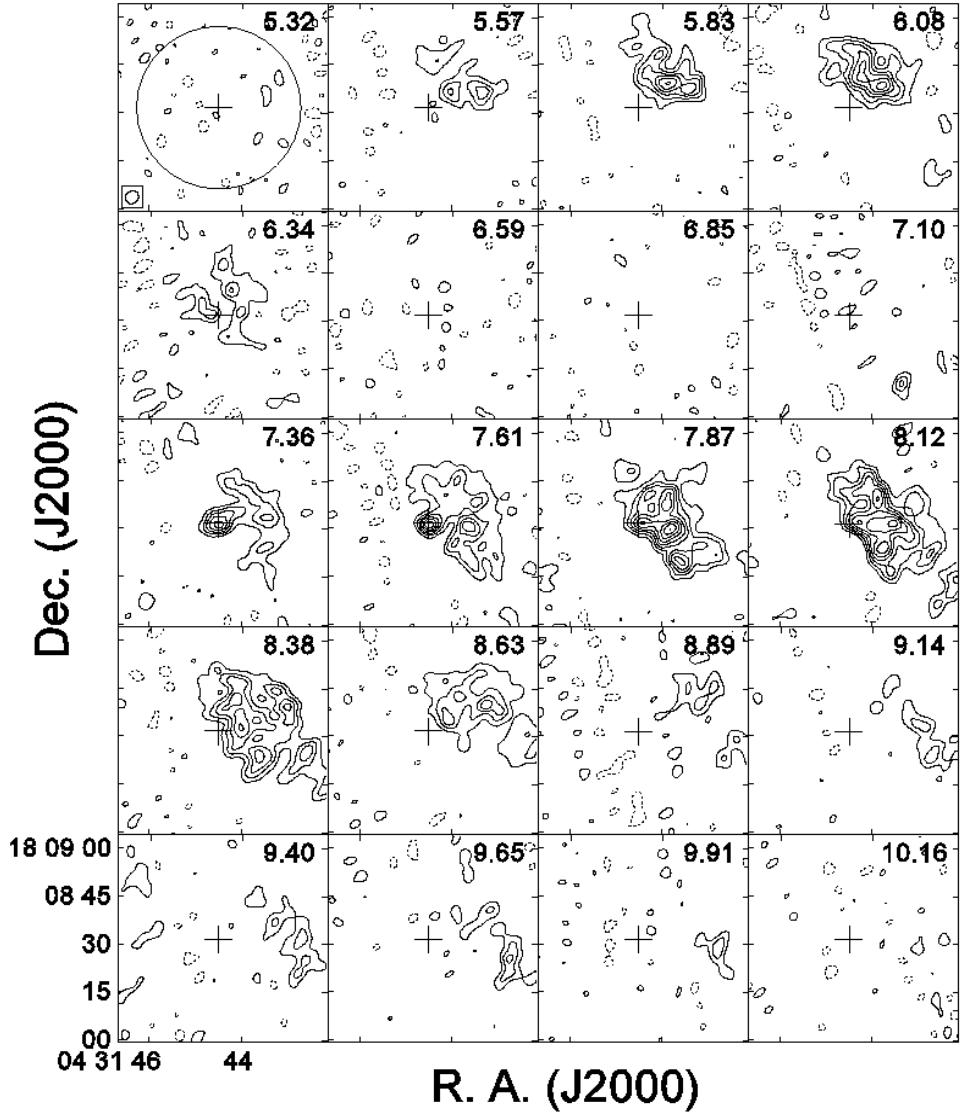}
\caption{
 Velocity channel maps of CS ($J=3-2$) emission. The
 central LSR velocity in unit of km s$^{-1}$ is denoted in the upper
 right corner of each panel. The contour lines are written in the same
 manner as in Figure \ref{fig:fig4} with 1~$\sigma = 0.18$ Jy
 beam$^{-1}$. 
 The central cross exhibits the position of NE. The open
 ellipse at the bottom-left corner of the top-left panel is the
 synthesized beam (HPBW).\label{fig:fig6}
 }
\end{figure}

\placefigure{fig:fig4}
\placefigure{fig:fig5}

Figure \ref{fig:fig5} shows CS ($J=2-1$) maps for the LVB
($V_{\rm{lsr}} = 5.32 - 6.47$ km s$^{-1}$), LVR ($V_{\rm{lsr}} = 7.04 -
8.57$ km s$^{-1}$), and HVR ($V_{\rm{lsr}} = 8.57 - 10.11$ km s$^{-1}$)
components.  
The LVB component is mainly distributed to the northwest of
NE and its total flux density integrated over the region above the
3~$\sigma$ level is 6.8 Jy km s$^{-1}$.   
The LVR component shows an inclined ``W'' shaped structure
with a southwest detached component. 
The overall emission is distributed along the north-south direction and
is shifted to the west of NE. 
One bright peak is the NE component, and the other bright peaks show an
arc-shaped structure open toward NE with a length of $\sim$ 7000 AU, as
shown by the dashed line in Figure \ref{fig:fig5}b. 
Although all the peaks seem to compose an arc open toward the west or
IRS 5 as previously suggested (e.g., Plambeck \& Snell 1995), we
consider that the NE peak corresponds to the circumbinary disk while the
other peaks, which originate from the gas swept up by the outflow from IRS 5,
make the arc open toward NE. This arc-like structure is seen in the
CS ($J=2-1$) maps at $V_{\rm{lsr}} = 7.34$ km s$^{-1}$ and 7.53 km
s$^{-1}$ in Figure \ref{fig:fig4}, and becomes more
prominent in the CS ($J=3-2$) maps at $V_{\rm{lsr}} = 7.36$ km s$^{-1}$
and 7.61 km s$^{-1}$ in Figure \ref{fig:fig6}.  
The arc can also be identified in the CS ($J=2-1$) maps from
$V_{\rm{lsr}} = 8.48$ km s$^{-1}$ to 10.01 km s$^{-1}$ in Figure
\ref{fig:fig4}: the peak seen in $V_{\rm{lsr}}$ $\ge$ 9.25 km s$^{-1}$
gradually splits into the two peaks at the lower red-shifted velocities,
which can be explained in terms of a bow shock around NE caused by the outflow
from IRS 5. 
The middle part of the arc seems relatively weak, as easily recognized
in the channel maps at $V_{\rm{lsr}}$ = 7.14 and 7.34 km s$^{-1}$ of
Figure \ref{fig:fig4}. 
This might be due to the blowing off of the dense gas by the collimated
outflows or the HH jets from NE at P.A. $\approx$ 243$^{\circ}$.     
The total flux density above the 3~$\sigma$ level is 21.9 Jy km s$^{-1}$.  
The HVR component is distributed on the west side of the
LVR component, and its total flux density is 8.1 Jy km s$^{-1}$
(see Table \ref{tbl:tbl3}).  

\begin{figure}
\epsscale{1.0}
\plotone{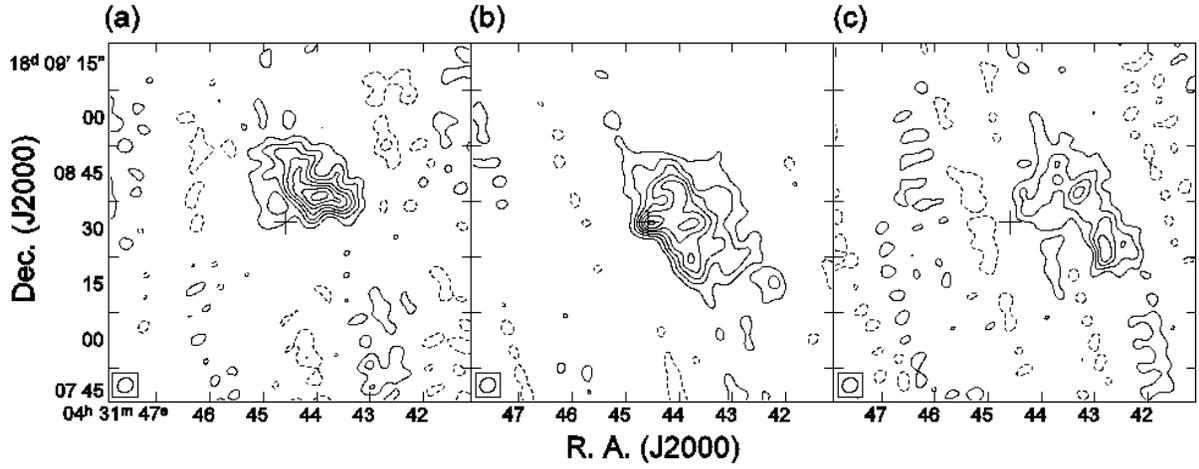}
\caption{
 CS ($J=3-2$) maps for (a) the LVB component of
 $V_{\rm{lsr}} = 5.43 - 6.47$ km s$^{-1}$, (b) the LVR component of
 $V_{\rm{lsr}} = 7.23 - 8.51$ km s$^{-1}$, 
 and (c) the HVR component of $V_{\rm{lsr}} = 8.51 -
 10.04$ km s$^{-1}$. The contour lines of each map are
 written in the same manner as in Figure \ref{fig:fig4}.
 The rms noise levels are 94 mJy beam$^{-1}$, 125 mJy
 beam$^{-1}$, and 75 mJy beam$^{-1}$ for (a), (b), and (c),
 respectively. The central cross is the position of NE. 
 \label{fig:fig7}} 
\end{figure}

\begin{figure}
\epsscale{1.0}
\plotone{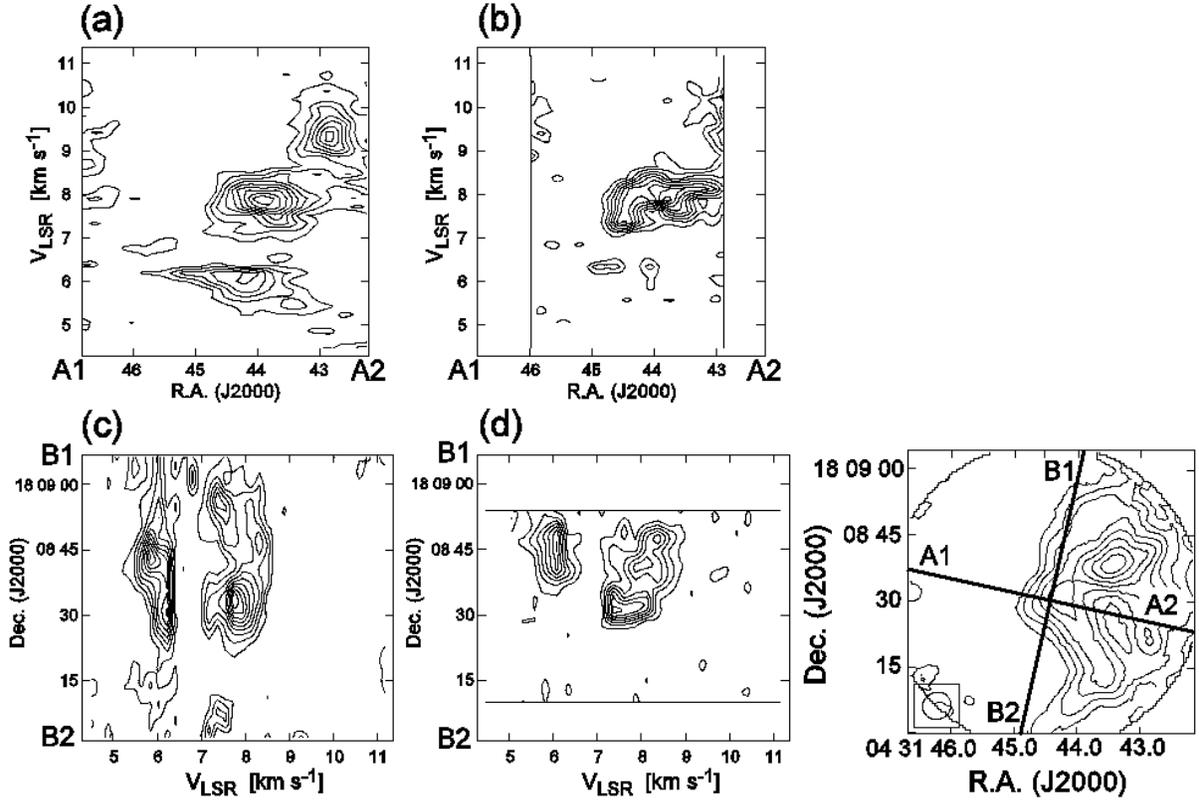}
\caption{
 (a) Position-velocity diagram of CS ($J=2-1$) emission cut along the
 line (A1 - A2) between L1551 NE and IRS 5, which is almost parallel to
 the outflow axis of IRS 5.  
 (b) P-V diagram of CS ($J=3-2$) emission along the line (A1 - A2).
 (c), (d) P-V diagrams of CS ($J=2-1$) and CS ($J=3-2$) emissions along
 the line (B1 - B2) perpendicular to the outflow axis.  
 The contour levels of all the diagrams are in steps of 10\% of the peak
 flux densities from 20\% to 100\%. \label{fig:fig8}}
\end{figure}

Figure \ref{fig:fig6} shows channel maps of CS ($J=3-2$)
with a velocity resolution of 0.25 km s$^{-1}$. The features of
the CS ($J=3-2$) emission shows a good agreement with those of CS
($J=2-1$) as follows: 
From $V_{\rm{lsr}} = 5.57$ km s$^{-1}$ to 6.34 km s$^{-1}$, extended
emission is seen to the north of NE. From $V_{\rm{lsr}} = 6.59$
km s$^{-1}$ to 7.10 km s$^{-1}$, the emission is absent because of the
resolving out. In the following channels from $V_{\rm{lsr}} = 7.36$ km
s$^{-1}$ to 8.12 km s$^{-1}$, particularly $V_{\rm{lsr}} = 7.36$ to
$7.61$ km s$^{-1}$, one can clearly see the bright NE
component surrounded by clumpy arc-shaped components which open toward NE. 
The size of the NE
component is 7$\farcs$4 $\times$ 4$\farcs$6 (FWHM) (1040 $\times$ 640
AU) at P.A. of 126$^\circ$.  
In $V_{\rm{lsr}} = 8.38$ km s$^{-1}$ to 9.91 km s$^{-1}$, clumpy emission
is seen to the west of the source. 
Figure \ref{fig:fig7} shows CS ($J=3-2$) maps for the LVB
($V_{\rm{lsr}} = 5.43 - 6.47$ km s$^{-1}$), LVR ($V_{\rm{lsr}} = 7.23 -
8.51$ km s$^{-1}$), and HVR ($V_{\rm{lsr}} = 8.51 - 10.04$ km s$^{-1}$)
components.  
The overall features are almost consistent with those of CS ($J=2-1$) in
Figure \ref{fig:fig5}, but the NE component in the LVR range is
much more prominent compared to the CS ($J=2-1$) case.  
This is probably because the NE component is denser and hotter than the
surrounding extended component. The total flux densities of the LVB, LVR,
and HVR components, are 22.9, 50.3, and 21.0 Jy km s$^{-1}$, respectively
(see Table \ref{tbl:tbl3}).   

\placefigure{fig:fig6}
\placefigure{fig:fig7}
\placetable{tbl:tbl3}

We estimate the total mass of the dense gas around NE. Although the
CS emission is generally optically thick around a protostar, the lower
limit of the mass can be obtained under the simple assumption of
optically thin condition. The mass of the dense gas can be calculated
by the following equation ~\citep{1986ApJ...303..416S}:  
\begin{eqnarray}
 M_{\rm{H_2}} = 4.6 \times 10^{-9} \frac{\exp(2.4/T_{\rm{ex}})}{1-\exp(-4.7/T_{\rm{ex}})}
\Biggl[\frac{10^{-9}}{X\rm{(CS)}}\Biggr] \nonumber \\
\left(\frac{d}{\rm{[pc]}}\right)^2 
\left(\frac{\lambda}{\rm{[mm]}}\right)^2 
\left(\frac{F_\nu}{\rm{[Jy km s^{-1}]}}\right)  M_\odot, \label{eq1}
\end{eqnarray} 
where $T_{\rm{ex}}$ is the excitation temperature, $X$(CS) is the
fractional abundance of CS relative to H$_2$, $d$ is the distance
to NE, and $F_\nu$ is the integrated flux density of CS ($J=2-1$) or CS
($J=3-2$).  
With $X$(CS) of 10$^{-9}$ ~\citep{1980ApJ...235..437L} and $T_{\rm{ex}}$ of
20 K ~\citep{1995ApJ...446..234P}, the total mass of the dense gas is 
estimated to be 0.18 $\pm$ 0.02 $M_\odot$, and the mass of the NE
component detected in the LVR range is estimated to be (1.0
$\pm$ 0.1) $\times$ 10$^{-2}$ $M_\odot$.    
The mass estimate is not sensitive to the adopted temperature: in the
case of $T_{\rm{ex}}$ = 50 K, the upper limit temperature
~\citep{1995ApJ...446..234P}, the total mass becomes 0.39 $\pm$ 0.04
$M_\odot$, and the NE component mass becomes (2.4 $\pm$ 0.3) $\times$
10$^{-2}$ $M_\odot$. Our total mass is in good agreement with the
previously reported mass of $0.016 - 0.02$ $M_\odot$ from CS
observations, considering $X$(CS) $= 10^{-8}$ they adopted
~\citep{1995ApJ...446..234P}.    
The mass of the NE component, however, is about a half of the mass
derived from the continuum observations, suggesting that the CS emission
is optically thick.    

Figure \ref{fig:fig8} shows P-V diagrams of the CS ($J=2-1$) and CS
($J=3-2$) emission of NE. In the diagram along the direction between
IRS 5 and NE (P.A. = 78$^{\circ}$, Figures \ref{fig:fig8}a \& b), 
which is roughly parallel to the outflow and jets of
IRS 5 ~\citep[P.A. $\approx$ 242$^\circ$]{1999AJ....118..972D}, there exist
three distinct components, corresponding to the LVB, LVR, and HVR components.
Note that the emission near the systemic velocity ($V_{\rm{lsr}}$
$\approx$ 6.7 km s$^{-1}$) is absent, owing to the resolving out of
extended components. Figure \ref{fig:fig8}a clearly shows a velocity
gradient along the direction between IRS 5 and NE: as farther from IRS 5
or closer to NE, the amount of redshift gets smaller.
This can be interpreted in terms of deceleration of the swept
up shell associated with the red-shifted outflow lobe of IRS 5.
In the diagram of CS ($J=2-1$) cut along the direction perpendicular to the
outflow axis of IRS 5 (Figure \ref{fig:fig8}c), on the other hand, no
systematic motion was detected. The P-V diagrams of CS ($J=3-2$) show
similar features to those of CS ($J=2-1$) (Figures \ref{fig:fig8}b \& d).  

\placefigure{fig:fig8}

\subsubsection{CS ($J=2-1$) emission with the Nobeyama 45 m
telescope}
\label{res2}
\begin{figure}
\epsscale{0.7}
\plotone{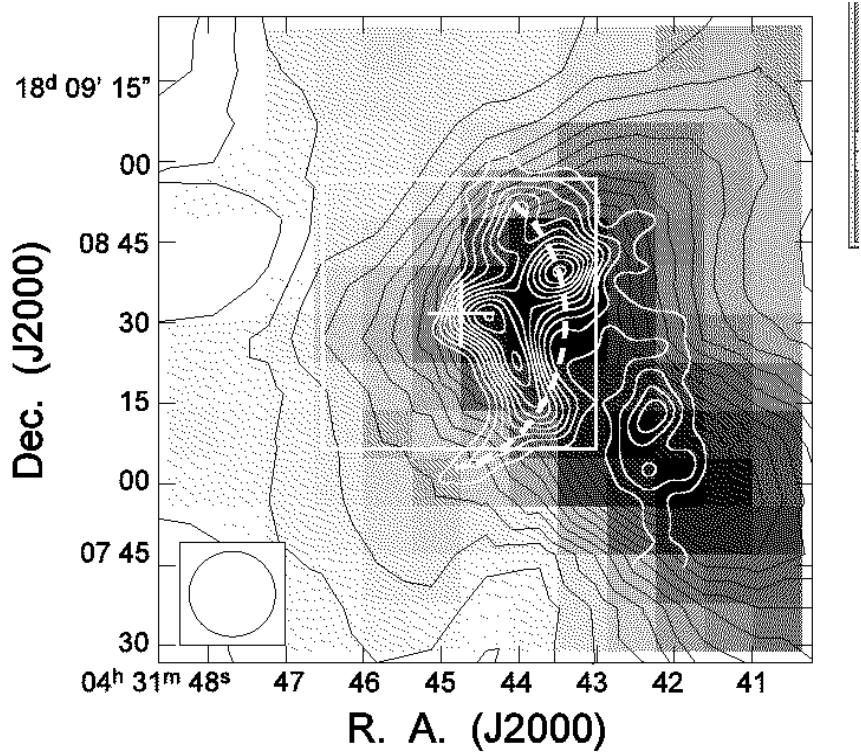}
\caption{
 Map of the LVR component of CS ($J=2-1$) emission
 ($V_{\rm{lsr}} = 7.04 - 8.57$ km s$^{-1}$) taken with the NMA
 superposed on a total integrated intensity map of CS ($J=2-1$)
 over $V_{\rm{lsr}} = 5.0 - 11.0$ km s$^{-1}$ taken with the Nobeyama 45
 m telescope. Black contour lines with the gray scale image are
 from the 45 m telescope in steps of $T_A^* = 0.2$ K.   
 Note the large scale CS arc-shaped structure open toward L1551 IRS 5 seen in
 the 45 m map and the small scale CS arc-shaped structure open toward L1551 NE
 seen in the NMA map which is denoted by a white dashed line. The white
 square shows an area of 50$\arcsec$ $\times$ 50$\arcsec$ around NE
 where the CS line profiles are calculated. The central cross indicates
 the position of NE. The open circle in the bottom-left corner is the
 angular resolution of the 45 m telescope (16$\arcsec$ in
 HPBW). \label{fig:fig9}}
\end{figure}

A total intensity map of CS ($J=2-1$) emission around NE integrated over
$V_{\rm{lsr}} = 5.0$ km s$^{-1}$ to 11.0 km s$^{-1}$, which is obtained
with the 45 m telescope, is shown as a gray scale image with contour
lines in Figure \ref{fig:fig9}.   
Two bright peaks can be seen: One is to the 15$\arcsec$ west of NE which
corresponds to the bright northwest peak seen in the NMA images of
Figures \ref{fig:fig2}a \& \ref{fig:fig5}b, and the other is to
the 40$\arcsec$ southwest of NE which corresponds to the southwest
detached component seen in Figure \ref{fig:fig5}b.  
The overall emission shows an arc-shaped structure (large-scale CS
arc-shaped structure) which is open to the
west, and NE is located at the eastern tip of the arc.  
The line profile integrated over a 50$\arcsec$ square around NE is shown
in Figure \ref{fig:fig3}. The profile is mainly red-shifted
compared to the systemic velocity of NE ($V_{\rm{lsr}}$ = 6.7 km
s$^{-1}$) where the self-absorption can be recognized. These features
have been found in previous studies (e.g., Plambeck \& Snell 1995), and
the large-scale CS arc-shaped structure has been thought to trace the
swept-up shell of the outflow powered by IRS 5.   
Another arc-shaped structure (small-scale CS arc-shaped structure)
indicated by the white dashed line in Figure \ref{fig:fig9} has been
newly discovered by this study (see also Figure \ref{fig:fig10}).

\placefigure{fig:fig9}

In order to estimate the missing flux of the NMA observations, 
we compare the CS ($J=2-1$) flux densities with the NMA and the 45 m
telescope in the same area of 50$\arcsec$ square around NE.
The total flux density with the 45 m telescope is 129 Jy km s$^{-1}$ in
the range $V_{\rm{lsr}} = 5.0$ km s$^{-1}$ to 11.0 km s$^{-1}$, while
that with the NMA is 36.8 Jy km s$^{-1}$ in almost the same velocity
range  $V_{\rm{lsr}} = 5.3 - 10.1 $km s$^{-1}$.
With the total flux density of the 45 m observations, equation
(\ref{eq1}) gives a total gas mass of 0.55 $\pm$ 0.07 $M_\odot$ under
the same assumptions as in the NMA case. The flux difference can mainly
be attributed to the resolving out of extended components in the NMA
observations.   
In the LVB ($V_{\rm{lsr}}$ $\lesssim 6.5$ km s$^{-1}$) and LVR
($V_{\rm{lsr}}$ $\approx 7.0 - 8.5$ km s$^{-1}$) 
ranges, the fraction of the missing flux density is about $60 - 70$ \%. 
In the HVR range ($V_{\rm{lsr}}$ $\gtrsim$ 8.5 km
s$^{-1}$) and around the systemic velocity ($V_{\rm{lsr}}$ $\sim$ 6.7 km
s$^{-1}$), more than 80 \% of the 45 m flux density is missing. 
These indicate that the smoothly extended components are dominant in the
HVR range as well as around the systemic velocity,
suggesting that the red-shifted outflow gas from IRS 5 has a smooth
structure.   

\section{Discussion} \label{bozomath}
The NMA observations of the CS ($J=2-1 /J=3-2$) emissions have revealed
the small-scale arc-shaped structure around NE and the
deceleration of the outflowing gas, suggesting that a distinct
interaction exists between NE and the outflowing gas.    
We discuss the interaction between them in \S~\ref{diss1}.  
Next, we propose a plausible scenario of the outflow-triggered formation
of NE in \S~\ref{diss2}. 
Finally, we discuss the possibility that the outflow from IRS 5 has
blown off the outer envelope around NE in \S~\ref{diss3}.

\subsection{Interaction between NE and the red-shifted outflow lobe of IRS 5}
\label{diss1} 
\begin{figure}
\epsscale{1.0}
\plotone{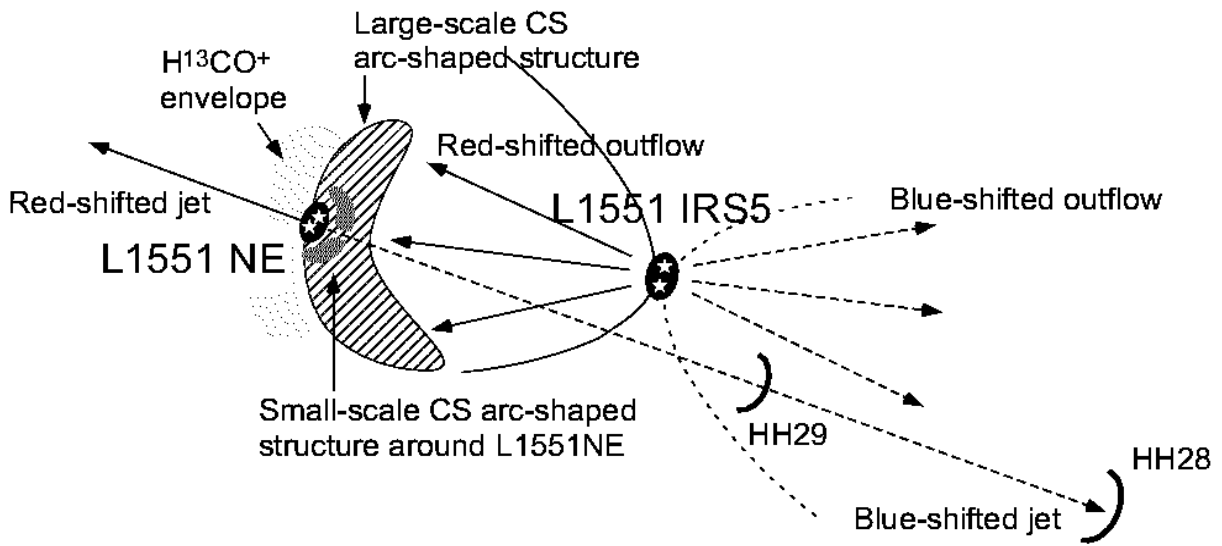}
\caption{
 Schematic illustration of the relationship between L1551 NE and
 IRS 5. The large-scale  CS arc-shaped structure open toward IRS 5
 represents the swept-up dense shell by the outflow from IRS 5. The
 small-scale CS arc-shaped structure open toward NE seems to be
 a bow-shock around NE caused by the impact of the dense shell. 
 The H$^{13}$CO$^{+}$ emission is distributed in the pre-shock region on
 the east side of NE, whereas the CS emission is mainly distributed in
 the post-shock region on the west side of NE. \label{fig:fig10}}
\end{figure}

The circumstellar environment around NE has been strongly affected by
the outflow from IRS 5, judging from the spatial distribution and velocity
structure of the CS emission around NE.  
Figure \ref{fig:fig10} is a schematic illustration of NE and IRS
5, which briefly summarizes our results. 
The large-scale CS arc-shaped structure open toward IRS 5 around NE, which
was revealed by the 45 m observations, is thought to trace the
swept-up shell of the outflow from IRS 5, because the shell exhibits the
highly red-shifted emission compared to the systemic velocities of IRS 5
and NE.  
Since NE is located at the leading edge of the shell, it is most likely
that the shell has impacted on the natal core of NE. 

The impact of the shell on NE is strongly supported by the small-scale
CS arc-shaped  structure, which is newly discovered by our observations.
The arc is open toward NE and exhibits the distinct deceleration feature 
along the outflow axis of IRS 5. These suggest that the swept-up 
gas accelerated by the outflow from IRS 5 has impacted on NE or 
the natal core of NE, forming the bow-shock-like arc as well as 
the deceleration revealed in Figure \ref{fig:fig8}a.  
Here, we estimate the deceleration of the gas, $a_{\rm{dcl}}$, under a
simple assumption of constant deceleration. 
The projected distance between NE and the peak position of the
HVR component in Figure \ref{fig:fig8}a is 20$\arcsec$,
corresponding to 2800 AU. 
The HVR component has $V_{\rm{lsr}}$ $\approx$ 9.2 km
s$^{-1}$, whereas the gas associated with NE has $V_{\rm{lsr}}$
$\approx$ 6.7 km s$^{-1}$.  
Considering the inclination angle of the outflow from IRS 5 to be $50 -
70^{\circ}$ with respect to the plane of the
sky~\citep{1986ApJ...304..459L,1994A&A...292..631F,1998ApJ...504..314M}, 
$a_{\rm{dcl}}$ is estimated to be $(0.44 - 1.9)$ $\times$ 10$^{-3}$ km
s$^{-1}$ yr$^{-1}$, and the timescale of the deceleration
becomes $(3.9 - 8.9)$ $\times$ 10$^3$ yr.

\placefigure{fig:fig10}

\begin{figure}
\epsscale{0.7}
\plotone{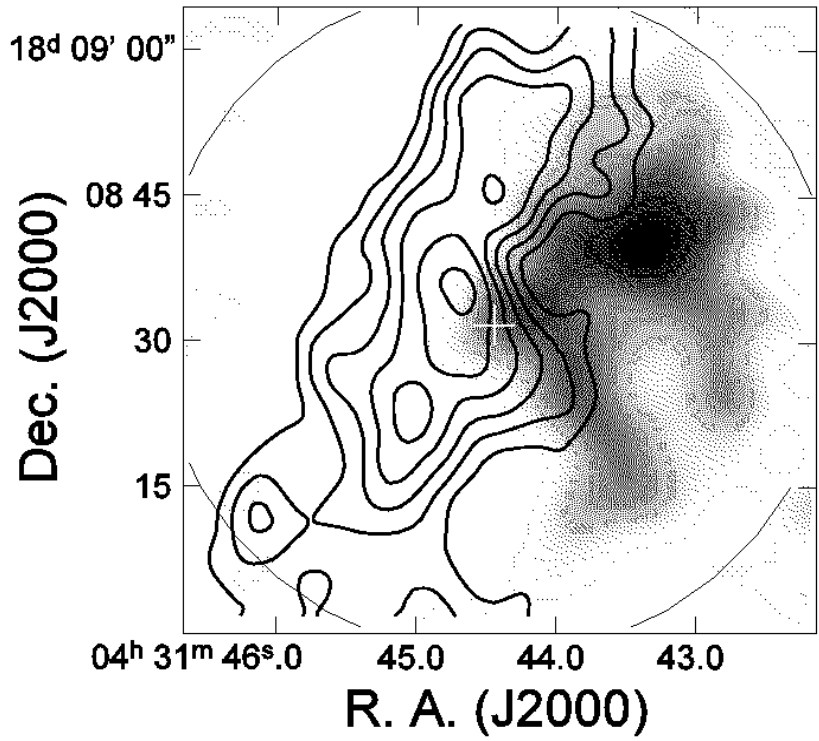}
\caption{
 Total intensity map of H$^{13}$CO$^+$ (solid contour)
 superposed onto the total intensity map of CS ($J=2-1$) in gray
 scale. The central cross indicates the position of NE.\label{fig:fig11}}
\end{figure}

The comparison of the CS and H$^{13}$CO$^+$ emissions
suggests that NE has formed at the shock front caused by the impact of
the CS shell.
Figure \ref{fig:fig11} shows the total intensity map of the
H$^{13}$CO$^+$ emission \citep{2001ApJ...547..840S} superposed onto the CS
($J=2-1$) map of NE by this study. Although both the CS and 
H$^{13}$CO$^+$ emissions trace dense gas, their distributions show clear
anti-correlation. The H$^{13}$CO$^+$ emission is mainly distributed to
the east of the source, which can be interpreted in terms of the depletion
of H$^{13}$CO$^+$ by the outflow ~\citep{2001ApJ...547..840S}:
Bachiller \& Perez Gutierrez (1997) suggested that the
H$^{13}$CO$^+$ molecules are rapidly destroyed in shocked regions owing
to dissociative recombination caused by enhanced electron 
abundance. Thus, the H$^{13}$CO$^+$ emission seems to be distributed in
the pre-shocked region. On the other hand, CS, a tracer of
dense and shocked gas, is mainly distributed to the west of NE,
suggesting that the CS emission is distributed in the post-shocked
region disturbed by the outflow of IRS 5. 
Therefore, it is most likely that NE is located at the
interface between the pre-shocked and post-shocked regions.        
This interpretation is also consistent with the fact that the
H$^{13}$CO$^+$ emission was only detected within a range of
$V_{\rm{lsr}} = 5.84 - 7.72$ km s$^{-1}$ and no high-velocity
components were detected, whereas the CS emission was detected in a wide
range of $V_{\rm{lsr}}$ $\approx$ $5.3 - 10.1$ km s$^{-1}$.

The timescale for the interaction between the natal core of NE 
and the outflow of IRS 5 is characterized by the crossing time in 
which the outflow passes through the core, although there is 
ambiguity in the core radius. The curvature radius of the small-scale 
CS arc-shaped structure, $4 \times 10^3$ AU, can be taken as the lower
limit for the core radius, giving an interaction timescale of $(1-2)
\times 10^4$ yr.
This timescale is only twice the timescale of the ongoing
deceleration of the CS gas of $(3.9-8.9) \times 10^3$ yr.  
The upper limit of the core radius, on the other hand, should be $\sim$
$10^4$ AU as  
suggested by the continuum map at 1.3 mm (Motte ~\& Andr{\'e} 2001), 
giving a timescale of $(2-5) \times 10^4$ yr. Note that the radius of
$10^4$ AU agrees with the mean radius of the cores in the TMC
\citep{2002ApJ...575..950O}. The possible range of the interaction
timescale, therefore, is $(1-5) \times 10^4$ yr. 
Since the separation between NE and IRS 5 is $2 \times 10^4$ AU, only
twice larger than the above upper limit of the core radius, the upper
limit timescale of the interaction can also be regarded as a possible
traveling time from IRS  5 to NE of the outflow ($(4-10) \times 10^4$
yr). 

\placefigure{fig:fig11}

\subsection{Outflow-triggered star formation}\label{diss2}
In addition to the strong interaction between NE and the outflow of IRS
5, we discuss the possibility that the formation of NE was triggered by
the impact of the outflow by comparing several timescales.
Motte ~\& Andr{\' e} (2001) estimated the age of NE to be  $(0.6 - 5.0)
\times 10^4$ yr: The shortest age was
derived from the assumption that the mass accretion rate declines
exponentially given by
\begin{equation}
\dot{M}_{\rm{acc}}(t) = \frac{M_{\rm{env}}(t)}{\tau} = \frac{M_0 \exp(-t/\tau)}{\tau}, 
\end{equation}
and the longest one was derived from the steady
mass accretion rate of 2.0 $\times$ 10$^{-6}$ $M_\odot$ yr$^{-1}$. 
The interaction timescale between the NE core and the outflow from IRS
5, on the other hand, is $(1-5) \times 10^4$ yr, and this agrees 
with the above-mentioned age of NE. Although both the estimates 
of timescale contain uncertainty, this agreement 
suggests that the mass accretion of NE was triggered by the 
outflow from IRS 5: it is unlikely that the NE formation and the 
interaction between NE and the outflow have occurred 
independently but simultaneously at the same place in such a 
short span compared with a period of the protostellar phase ($\sim 10^5$
yr). This would be the first evidence for the outflow triggered 
star formation in low-mass star forming regions. 

In the blueshifted outflow lobe of IRS 5, there is another CS 
clump which resembles the shell near NE in spatial distribution, 
sizescale and the amount of velocity shifts with respect to the systemic
velocity of IRS 5 (see Fig.3 of Plambeck \&
Snell 1995). A straight line connecting these two shells or clumps 
passes directly through IRS 5, suggesting these two shells were 
simultaneously ejected from IRS 5 with similar velocities. 
The separation from IRS 5 of the blueshifted shell, however, is 
about twice larger than that of the redshifted shell near NE. 
This difference might be caused by the obstruction by the NE core 
in the redshifted outflow lobe of IRS 5. 
If this is the case, the duration of the 
interaction between the NE core and the outflow from IRS 5 should be
comparable to the traveling time from IRS 5 to NE of the outflow. 

\subsection{Blown off envelope around L1551 NE}\label{diss3}
\begin{figure}
\epsscale{0.7}
\plotone{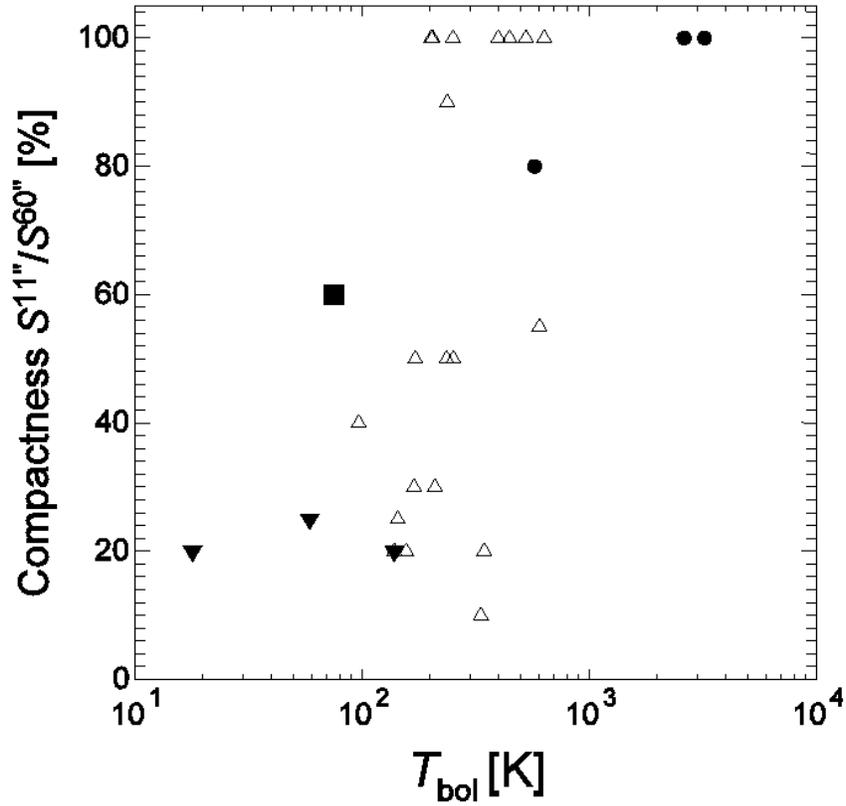}
\caption{
 Compactness of 1.3 mm continuum emission of YSOs in the TMC. The vertical
 axis shows the ratio of the peak flux density in an 11$\arcsec$ beam to
 the total flux density in a 60$\arcsec$ diameter circle. All the data are
 compiled from the survey with the IRAM 30 m telescope
 ~\citep{2001A&A...365..440M}. The filled triangles show class 0 objects,
 the open triangles show class I objects, and the filled circles show class II
 objects. The filled square indicates L1551 NE. \label{fig:fig12}}
\end{figure}

Class 0 protostars are usually surrounded by dense envelopes with a few
$\times$ 1000 AU sizes. As they evolve, the dense envelopes are
dispersed, and T Tauri stars without the envelopes appear
~\citep{2000ApJ...529..477L, 2001ApJ...547..840S}. 
NE, however, might be an exceptional class 0 protostar with a tiny
envelope, which is probably caused by the impact of the swept-up shell
of IRS 5. 
The compactness of the NE envelope is clearly demonstrated by the 1.3 mm
continuum imaging survey of YSOs with the IRAM 30 m telescope
~\citep{2001A&A...365..440M}. 
Figure \ref{fig:fig12} shows the compactness of YSOs defined by
the ratio of the peak intensity with an 11$\arcsec$ beam to the flux
density within a diameter of 60$\arcsec$ at 1.3 mm.
Although NE has a low bolometric temperature of 75 K, the degree of the
compactness of 60 \% is considerably high compared to the other class 0
objects.  
Moriarty-Schieven et al. (2000) also showed that the 1.3 mm flux
density of 0.851 $\pm$ 0.084 Jy obtained by the OVRO interferometer
(1$\farcs$29 $\times$ 1$\farcs$07 in HPBW) agrees with that of 0.83
$\pm$ 0.03 Jy obtained by the JCMT (20$\arcsec$ in HPBW) within the errors,
indicating that the dust emission is centrally condensed.  
Furthermore, the NE component seen in Figures \ref{fig:fig4} \&
\ref{fig:fig6} has a small size of $\sim$ 1000 AU, suggesting
the gas envelope is also compact. 
One possible interpretation of such an unusually compact
envelope is that the swept-up shell accelerated by the outflow from IRS 5,
which likely to trigger the NE formation, has blown off the outer parts
of the envelope.

\placefigure{fig:fig12}

The outflow of IRS 5 seems to have enough power to blow off the
circumstellar envelope around NE as follows:
The momentum flux put into the dense gas around NE by the decelerating
swept-up shell is given by 
\begin{equation}
P_{\rm{shell}} = M_{\rm{H_2}} \times a_{\rm{dcl}},
\end{equation}
where $M_{\rm{H_2}}$ is the mass of the swept-up shell of 0.18
$M_\odot$, and is estimated to be $(0.79 - 3.4) \times$ 10$^{-4}$
$M_\odot$ km s$^{-1}$ yr$^{-1}$. 
Here we assume that all the momentum flux which the outflow lost is put
into the dense gas for simplicity.
On the other hand, for a typical protostar with a mass of 0.5 $M_\odot$
surrounded by an envelope with 2000 AU radius and 0.1$ M_\odot$ mass,
the momentum required to  dissipate the envelope would be 7.0 $\times$
10$^{-2}$ $M_\odot$ km s$^{-1}$. 
Since the age of NE is $(0.6 - 5) \times$ 10$^4$ yr, the
momentum flux required to blow off the envelope have to be larger than
$(0.14 - 1.2) \times$ 10$^{-5}$ $M_\odot$ km s$^{-1}$ yr$^{-1}$.  
This flux is one order of magnitude smaller than $P_{\rm{shell}}$, indicating that
the outflow from IRS 5 might have blown off the outer parts of the
envelope around NE.     

\section{Summary}
We have presented the new results of the high angular resolution and high 
sensitivity CS ($J=2-1$/$J=3-2$) observations of the protostar L1551
NE. The main results are as follows: 
\begin{enumerate}
 \item{
      Both the CS ($J=2-1$) and ($J=3-2$) emissions were detected from
      $V_{\rm{lsr}}$ $\approx$ 5.3 to 10.1 km s$^{-1}$, and their
      distributions are quite similar to each other. 
      The arc-shaped structure open toward NE and the compact disklike
      structure at the NE position were clearly resolved. The total mass
      of the dense gas is estimated to be 0.18 $\pm$ 0.02 $M_\odot$, and
      the mass contained in the disklike structure is (1.0 $\pm$ 0.1)
      $\times$ 10$^{-2}$ $M_\odot$. 
      }  
 \item{
      One-arcsec resolution imaging of 3 mm and 2 mm continuum emission
      has revealed the disklike structure, perpendicular to the
      [Fe II] jet and the HH jets HH 28, 29, and 454. 
      The size of the structure is
      1$\farcs$39 $\times$ 1$\farcs$00 (200 AU $\times$ 140 AU) at
      P.A. of 159 $\pm$ 13$^\circ$ and the estimated mass is 0.032 $\pm$
      0.003 $M_\odot$. Although NE is a binary system consisting of NE-A
      and NE-B, the main contributor of the dust continuum
      emission seems to be NE-A.
      }
 \item{
      The CS gas around NE shows the distinct velocity gradient along
      the outflow axis of IRS 5. This can be explained by the
      deceleration of the swept-up shell accelerated by the outflow from
      IRS 5, suggesting that the swept-up shell has impacted on NE.  
      }
 \item{
      The distributions of the CS and H$^{13}$CO$^+$ emissions show clear
      anti-correlation, suggesting that CS is abundant in the
      post-shocked regions, whereas H$^{13}$CO$^+$ is abundant in the
      pre-shocked regions. 
      }
 \item{
      Since the timescale of the interaction is comparable to the
      age of NE, it is likely that the formation of NE was triggered by
      the impact of the swept-up shell of the outflow from IRS 5. 
      }
 \item{
      Our and previous results suggest that NE has a compact circumstellar
      envelope with a size of $\sim$ 1000 AU, although NE is categorized as
      a class 0 object. The outer parts of the envelope might be blown
      off by the outflow from IRS 5, because the momentum put in by the
      swept-up shell is sufficient to blow off the envelope.   
      } 
\end{enumerate}

Acknowledgements\\
We are grateful to the staff of the Nobeyama Radio Observatory (NRO)
for both operating the Millimeter Array and helping us with data
reduction. We also thank an anonymous referee for providing helpful
suggestions to improve the paper. S. Y. was financially supported by a
Research Fellowship of the Japan Society for the Promotion of Science
for Young Scientists. 





\clearpage



\clearpage

\clearpage

\begin{deluxetable}{lcc}
\tabletypesize{\footnotesize} 
\tablewidth{0pc} 
\tablecaption{Observational parameters \label{tbl:tbl1}} 
\tablehead{ 
\colhead{Observation period}& \multicolumn{2}{c}{1999 November - 2002
 February}}
\startdata
 Molecular line        & CS ($J=2-1$) & CS ($J=3-2$) \\
 \cline{1-3}
 Frequency [GHz]       &  97.980968   & 146.969049  \\
 Velocity resolution [km s$^{-1}$] & 0.19 &0.25  \\
 Synthesized beam size  & 6$\farcs$53 $\times$ 6$\farcs$15 &
 4$\farcs$35 $\times$ 3$\farcs$86 \\
 Position angle [$^{\circ}$] &  -56.0   &  -44.9 \\
 Noise level (1~$\sigma$) [Jy beam$^{-1}$] & 0.12 & 0.18 \\
 &&\\
 \cline{1-3}\\
 Continuum [GHz] & 98 $\pm$ 0.512, 110 $\pm$ 0.512& 135 $\pm$ 0.512, 147 $\pm$ 0.512 \\
 \cline{1-3}
 Synthesized beam size (D/AB)$^*$  & 6$\farcs$28 $\times$ 5$\farcs$87 /
 1$\farcs$52 $\times$ 1$\farcs$35 &
 4$\farcs$51 $\times$ 4$\farcs$19 / 1$\farcs$18 $\times$ 1$\farcs$10\\
 Position angle [$^{\circ}$] (D/AB)$^*$ & -49.9/ 5.2   & -41.6/ 54.8  \\
 Noise level (D/AB)$^*$ [mJy beam$^{-1}$] & 1.4 / 2.0 & 2.6 / 2.6 \\
 &&\\
 \cline{1-3}
 Phase $\&$ amplitude calibrator (flux density) &     & 0446+112 (1.4Jy)\\
 & 0507+179 (1.4Jy) & 0507+179 (1.9Jy)\\
 &                  & 0528+134 (2.6Jy)\\
 Bandpass calibrator &  \multicolumn{2}{c}{3C454.3, 3C279} \\
 $T_{\rm{sys}}$ [K] (in DSB)        & 150 K & 200 K  \\ 
\enddata 
\tablecomments{$^*$ D and AB are the NMA array configurations D (lowest angular
 resolution) and AB (highest angular resolution), respectively. }
\end{deluxetable}

\begin{deluxetable}{cccccccc}
\tabletypesize{\scriptsize} 
\tablewidth{0pc} 
\tablecaption{Total flux densities of the continuum emission of L1551 NE
\label{tbl:tbl2}} 
\tablehead{ 
 \colhead{Frequency}&\colhead{Total flux density}&
 \colhead{Peak flux density}&\colhead{Deconvolved size}&
 \colhead{P.A.}&\colhead{Beam size}&\colhead{P.A.}&\colhead{NMA
 config.}\\
 \colhead{}&\colhead{[mJy]}&
 \colhead{[mJy beam$^{-1}$]}&\colhead{[ $\arcsec$ ]}&
 \colhead{[ $^{\circ}$ ] }&\colhead{[ $\arcsec$ ]}&\colhead{[ $^{\circ}$ ]}&\colhead{}}
\startdata
104 GHz &  92.5 $\pm$ 9.4 &  74.6 &3.99 $\times$ 2.64 & 147 &6.28 $\times$ 5.87 &-49.9 &  D \\
        &  68.3 $\pm$ 7.1 &  41.5 &1.36 $\times$ 0.78 & 154 &1.52
 $\times$ 1.35 &  5.2 &  AB 
\\
\cline{1-8}
&&&&&&&\\
141 GHz & 202 $\pm$ 22 & 159 & 2.41   $\times$ 1.12 & 155 &4.51 $\times$ 4.19 & -41.6 & D\\ 
        & 130 $\pm$ 13 &  60.4 & 1.43 $\times$ 1.29 & 180 &1.18 $\times$ 1.10 & 54.8 & AB\\
\enddata 
\end{deluxetable} 

\begin{deluxetable}{lccccccc}
\tabletypesize{\footnotesize} 
\tablewidth{0pc} 
\tablecaption{CS ($J=2-1$) and CS ($J=3-2$) flux densities of L1551 NE
\label{tbl:tbl3}} 
\tablehead{ 
 \colhead{}&
 \multicolumn{2}{l}{CS ($\it{J}$=2-1)}&\colhead{}&
 \multicolumn{2}{l}{CS ($\it{J}$=3-2)}&\colhead{}&
 \colhead{}}
\startdata
&$V_{\rm{lsr}}$ & $\int S_\nu dv$ && $V_{\rm{lsr}}$ & $\int S_\nu dv$ &&Mean Mass \\
& [ km s$^{-1}$ ] & [ Jy kms$^{-1}$ ] && [ km s$^{-1}$ ]& [ Jy
 kms$^{-1}$ ]&& [10$^{-2}$ $\times$ $M_\odot$] \\
 \cline{1-8}
Low-velocity blue (LVB)&  5.32 - 6.47   & 6.8&& 5.43 - 6.47  & 22.9&& 3.9 $\pm$ 0.5 \\
NE component      &  6.47 - 8.57   & 2.6&& 6.47 - 8.51  &  4.9&& 1.0 $\pm$ 0.1 \\
Low-velocity red (LVR)&  7.04 - 8.57   &21.9&& 7.23 - 8.51  & 50.3&& 10.0$\pm$ 0.8 \\
High-velocity red  (HVR)&  8.57 - 10.11  & 8.1&& 8.51 - 10.04 & 21.0&& 4.0 $\pm$ 0.4 \\
\cline{1-8}
 Total            &              &36.8&&           & 94.2&&17.9 $\pm$ 1.7 
\enddata 
\end{deluxetable} 
 


\begin{thebibliography}{}
  \bibitem[Bachiller et al.(1994)Bachiller, Tafalla, \& Cernicharo]{1994ApJ...425L..93B} 
						   Bachiller, R., Tafalla, M., \& Cernicharo, J.\ 1994, \apjl, 425, L93 
						   
  \bibitem[Bachiller \& Perez Gutierrez(1997)]{1997ApJ...487L..93B} 
						   Bachiller, R.~\& Perez Gutierrez, M.\ 1997, \apjl, 487, L93 
						   
  \bibitem[Barsony \& Chandler(1993)]{1993ApJ...406L..71B} Barsony, M.~\& Chandler, C.~J.\ 1993, \apjl, 406, L71 
						   
  \bibitem[Beckwith et al.(1990)]{1990AJ.....99..924B} 
						   Beckwith, S.~V.~W., Sargent, A.~I., Chini, R.~S., \& Guesten, R.\ 1990, 
						   \aj, 99, 924 
						   
  \bibitem[Blaauw(1991)]{1991psfe.conf..125B} Blaauw, A.\ 1991, NATO ASIC 
						   Proc.~342: The Physics of Star Formation and Early Stellar Evolution, 125 
						   
  \bibitem[Devine et al. (1999)Devine, Reipurth, \& Bally]{1999AJ....118..972D} Devine, D., 
						   Reipurth, B., \& Bally, J.\ 1999, \aj, 118, 972
						   
  \bibitem[Emerson et al.(1984)]{1984ApJ...278L..49E} Emerson, J.~P., Harris, 
						   S., Jennings, R.~E., Beichman, C.~A., Baud, B., Beintema, D.~A., Wesselius, 
						   P.~R., \& Marsden, P.~L.\ 1984, \apjl, 278, L49 
						   
  \bibitem[Elias(1978)]{Elias78}Elias, J. H. 1978. \apj, 224, 857 
						   
  \bibitem[Fridlund \& Liseau(1994)]{1994A&A...292..631F} Fridlund, 
						   C.~V.~M.~\& Liseau, R.\ 1994, \aap, 292, 631 
						   
  \bibitem[Gomez et al.(1993)]{1993AJ....105.1927G} 
						   Gomez, M., Hartmann, L., Kenyon, S.~J., \& Hewett, R.\ 1993, \aj, 105, 1927 
						   
  \bibitem[Linke \& Goldsmith(1980)]{1980ApJ...235..437L} Linke, R.~A.~\& Goldsmith, P.~F.\ 1980, \apj, 235, 437 
						   
  \bibitem[Liseau \& Sandell(1986)]{1986ApJ...304..459L} Liseau, R.~\& Sandell, G.\ 1986, \apj, 304, 459 
						   
  \bibitem[Looney et al.(2000)Looney, Mundy, \& Welch]{2000ApJ...529..477L} Looney, L.~W., 
						   Mundy, L.~G., \& Welch, W.~J.\ 2000, \apj, 529, 477 

  \bibitem[Momose et al.(1998)]{1998ApJ...504..314M} Momose, M., Ohashi, N., 
						   Kawabe, R., Nakano, T., \& Hayashi, M.\ 1998, \apj, 504, 314 
						   
  \bibitem[Moriarty-Schieven et al.(1994)]{1994ApJ...436..800M} Moriarty-Schieven, G.~H., Wannier, 
						   P.~G., Keene, J., \& Tamura, M.\ 1994, \apj, 436, 800 
						   
  \bibitem[Moriarty-Schieven et al.(2000)]{2000ApJ...533L.143M} 
						   Moriarty-Schieven, G.~H., Powers, J.~A., Butner, H.~M., Wannier, P.~G., \& Keene, J.\ 2000, \apjl, 533, L143 
						   
  \bibitem[Motte \& Andr{\' e}(2001)]{2001A&A...365..440M} Motte, F.~\& Andr{\' e}, P.\ 2001, \aap, 365, 440 
						   
  \bibitem[Okumura et al.(2000)]{2000PASJ...52..393O} Okumura, S.~K.~et al.\ 
						   2000, \pasj, 52, 393 
						   
  \bibitem[Onishi et al.(2002)]{2002ApJ...575..950O} Onishi, T., Mizuno, A., 
								   Kawamura, A., Tachihara, K., \& Fukui, Y.\ 2002, \apj, 575, 950 

  \bibitem[Plambeck \& Snell(1995)]{1995ApJ...446..234P} Plambeck, R.~L.~\& Snell, R.~L.\ 1995, \apj, 446, 234 
						   
  \bibitem[Reipurth et al.(1997)Reipurth, Bally, \& Devine]{1997AJ....114.2708R} Reipurth, 
						   B., Bally, J., \& Devine, D.\ 1997, \aj, 114, 2708 
						   
  \bibitem[Reipurth et al.(2000)]{2000AJ....120.1449R} Reipurth, B., Yu, K., 
						   Heathcote, S., Bally, J., \& Rodr{\' i}guez, L.~F.\ 2000, \aj, 120, 1449 
						   
  \bibitem[Reipurth \& Bally(2001)]{2001ARA&A..39..403R} Reipurth, B.~\&
						   Bally, J.\ 2001, \araa, 39, 403  
						   
  \bibitem[Reipurth et al.(2002)]{2002AJ....124.1045R} Reipurth, B., Rodr{\' i}guez, L.~F., 
						   Anglada, G., \& Bally, J.\ 2002, \aj, 124, 1045 
						   
  \bibitem[Saito et al.(2001)]{2001ApJ...547..840S} 
						   Saito, M., Kawabe, R., Kitamura, Y., \& Sunada, K.\ 2001, \apj, 547, 840 
						   
  \bibitem[Scoville et al.(1986)]{1986ApJ...303..416S} Scoville, N.~Z., 
								   Sargent, A.~I., Sanders, D.~B., Claussen, M.~J., Masson, C.~R., Lo, K.~Y., 
								   \& Phillips, T.~G.\
								   1986,
								   \apj,
								   303,
								   416
								   
 \end{thebibliography}
\end{document}